\documentclass[%
 reprint,superscriptaddress,amsmath,amssymb,
 aps,nofootinbib]{revtex4-1}
\usepackage{dsfont}
\usepackage{graphicx}
\usepackage{dcolumn}
\usepackage{bm}
\usepackage{color}
\usepackage{xcolor}
\usepackage{epsfig}
\usepackage{soul}
\usepackage[colorlinks=true,urlcolor=brown,linkcolor=blue,citecolor=magenta]{hyperref}
\usepackage{natbib}
\usepackage{float}
\usepackage{footmisc}
\definecolor{lightred}{rgb}{1,0.5,0.5}
\definecolor{lightgreen}{rgb}{0.5,1,0.5}
\definecolor{lightblue}{rgb}{0.5,0.5,1}
\definecolor{lightcyan}{rgb}{0.5,0.75,0.75}
\definecolor{lightmagenta}{rgb}{0.75,0.5,0.75}
\definecolor{customgreen}{rgb}{0.494,1,0.502}

\newcommand{\keV}{\mathinner{\mathrm{keV}}}
\newcommand{\MeV}{\mathinner{\mathrm{MeV}}}
\newcommand{\GeV}{\mathinner{\mathrm{GeV}}}
\newcommand{\TeV}{\mathinner{\mathrm{TeV}}}

\begin{document}

\begin{flushright}
  DESY-22-157 
\end{flushright}

\title{Electroweak Asymmetric Early Universe via a Scalar Condensate}

\author{Jae Hyeok Chang} 
\email{jaechang@umd.edu}
\affiliation{Department of Physics and Astronomy, Johns Hopkins University, Baltimore, MD 21218, USA}
\affiliation{Maryland Center for Fundamental Physics, Department of Physics, University of Maryland, College Park, MD 20742, USA}

\author{María Olalla Olea-Romacho }
\email{mariaolalla.olearomacho@phys.ens.fr}
\affiliation{Laboratoire de Physique de l'École Normale Supérieure, ENS, Université PSL, CNRS, Sorbonne Université, Université Paris Cité, F-75005 Paris, France}
\affiliation{Deutsches Elektronen-Synchrotron DESY, Notkestr.
85, 22607 Hamburg, Germany}

\author{Erwin H. Tanin}
\email{etanin1@jhu.edu}
\affiliation{Department of Physics and Astronomy, Johns Hopkins University, Baltimore, MD 21218, USA}

\begin{abstract}
    Finite temperature effects in the Standard Model tend to restore the electroweak symmetry in the early universe, but new fields coupled to the higgs field may as well reverse this tendency, leading to the so-called electroweak symmetry non-restoration (EW SNR) scenario. Previous works on EW SNR often assume that the reversal is due to the thermal fluctuations of new fields with negative quartic couplings to the higgs, and they tend to find that a large number of new fields are required. We observe that EW SNR can be minimally realized if the field(s) coupled to the higgs field develop(s) a stable condensate. We show that one complex scalar field with a sufficiently large global-charge asymmetry can develop a condensate as an outcome of thermalization and keep the electroweak symmetry broken up to temperatures well above the electroweak scale. In addition to providing a minimal benchmark model, our work hints on a class of models involving scalar condensates that yield electroweak symmetry non-restoration in the early universe.
\end{abstract}

\maketitle
\section{Introduction}

It is an empirical fact that we live at present in a vacuum that breaks the electroweak (EW) symmetry. At high temperatures, the higgs field acquires positive thermal mass squared contributions from the fermions and gauge bosons coupled to it. This thermal mass tends to confine the higgs 
field at the origin, leading to the standard theoretical expectation that the EW symmetry is restored in the early universe. The latter scenario is true within the Standard Model (SM) and assumed in most beyond the Standard Model explorations in the literature. However, thus far there has been no evidence for a period of restored EW symmetry in the early universe and current observational limits permit a wide variety of extensions to the SM which might reverse the tendency to restore the EW symmetry at high-temperatures.\footnote{The possibility of high-temperature symmetry non-restoration, not specific to the EW sector, was first studied in \cite{Weinberg:1974hy,Mohapatra:1979qt,Mohapatra:1979vr,Langacker:1980kd,Salomonson:1984rh,Dvali:1995cc,Dvali:1995cj,Bajc:1997ky}.} Indeed, counterexamples to the conventional picture presented above do exist. Such alternative scenarios 
where the EW symmetry remains broken at temperatures above the EW scale feature the phenomenon commonly referred to as electroweak symmetry non-restoration (EW SNR).

The possibility of the higgs field acquiring a nonzero vacuum expectation value (vev)
in the early universe has wide reaching consequences, some of which have been explored in~\cite{Yang:2015ida,Freese:2017ace,Passaglia:2021upk,Meade:2018saz,Baldes:2018nel,Glioti:2018roy, Espinosa:2004pn, PhysRevD.97.123509,Matsedonskyi:2020kuy, Matsedonskyi:2020mlz, Bai:2021hfb, Biekotter:2021ysx, Carena:2021onl,Matsedonskyi:2021hti}. The electroweak phase transition may not have occurred or instead took place at a much higher temperature as compared to the SM prediction. Sphaleron processes would remain suppressed at temperatures well above the electroweak scale, thus making e.g.~high-temperature electroweak baryogenesis viable. Early universe calculations that rely on the properties of the primordial SM plasma would have to be appropriately modified. The impacts of these modifications may be imprinted in relics such as gravitational waves, dark matter, and dark radiation that decoupled early. It is therefore important to consider the less explored possibility that the broken electroweak phase persists above the electroweak scale.

One way to modify the thermal evolution of the higgs vev is to couple the higgs field to new scalar degrees of freedom via higgs-portal couplings. If these quartic couplings are negative,\footnote{These quartic couplings are defined as negative in the potential and positive in the Lagrangian.} the higgs field would acquire a negative thermal mass from the thermal fluctuations of the new scalars. This fact was utilized to achieve EW SNR in \cite{Meade:2018saz,Baldes:2018nel,Glioti:2018roy}. There it was shown that the presence of at least $O(100)$ thermalized scalars with negative quartic couplings to the higgs field can keep the electroweak symmetry broken at temperatures well above the electroweak scale. Further studies revealed a variety of models that display electroweak symmetry non-restoration \cite{Matsedonskyi:2020mlz,Biekotter:2021ysx, Carena:2021onl,Matsedonskyi:2021hti, Bai:2021hfb, Biekotter:2022kgf}, and yet the presence of a large number of new degrees of freedom remains to be a common feature of existing models for this phenomenon. In some of these models~\cite{Matsedonskyi:2021hti,Biekotter:2021ysx}, EW SNR can technically be realized with an $O(1)$ number of new fields at the cost of limiting the highest temperature to which EW SNR can be reliably sustained, which in both studies did not go beyond $\sim 100 \TeV$.

While increasing the multiplicity of the fields coupled to the higgs field helps to alleviate various constraints~\cite{Meade:2018saz,Baldes:2018nel,Glioti:2018roy}, in particular those related to the stability of the scalar potential and perturbativity, this feature is not a proximate cause of the high-temperature EW SNR phenomenon. In this paper, we show that the addition of one scalar field coupled to the higgs is sufficient to achieve EW SNR if the scalar develops a sufficiently large vev at high temperatures in the early universe. Note that increasing the scalar vev neither destabilizes the scalar potential nor exacerbates the running of couplings. We demonstrate this idea of realizing EW SNR via a vev in a simple model of a complex scalar singlet coupled to the EW sector through the higgs portal with negative coupling. In the presence of a sufficiently large chemical potential, the thermal equilibrium state of the new scalar includes a Bose-Einstein condensate (BEC) \cite{Haber:1981ts, Mangano:2001ix} and this condensate yields the requisite large negative higgs mass squared for EW SNR.

Chemical potentials in the universe naturally arise in the presence of net background charges associated with some global symmetries. In fact, current observations are consistent with the universe possessing large background charges of certain kinds. While the baryon asymmetry of the universe has been observed to be tiny, $n_B/s\sim 10^{-10}$ \cite{Cyburt:2015mya,Planck:2018vyg}, up to $O(1)$ total lepton asymmetry~\cite{Barenboim:2017dfq} is still allowed. Charge asymmetries may also reside in the dark sector~\cite{Kaplan:2009ag} at an unconstrained level. Global symmetries are expected to be broken at high energies by higher dimensional operators \cite{Kamionkowski:1992mf,Holman:1992us}. Thus, a field whose Lagrangian respects a global symmetry at low energies could carry a net charge as an after effect of its high-energy dynamics. A concrete example of this is the Affleck-Dine mechanism~\cite{Affleck:1984fy}. Furthermore, if some form of entropy production \cite{Randall:2015xza,Bramante:2017obj,Evans:2019jcs} or charge washout \cite{Buckley:2011ye, Cirelli:2011ac,Tulin:2012re} took place, these charge asymmetries could be much greater in the early universe and have stronger impacts then.

In this paper we show that EW SNR can be minimally realized by coupling the higgs to a scalar that develops a vev in the early universe. We elaborate this point further in section~\ref{s:II}. In section~\ref{s:III}, we present a simple example model (with a new complex scalar that forms a BEC) that demonstrates this idea, analyze the viable parameter space for achieving EW SNR, and describe its cosmology. Finally, we conclude in section~\ref{s:IV}.

\section{High-Temperature Electroweak Symmetry Non-restoration with a Scalar Condensate}
\label{s:II}

The higgs doublet $\mathcal{H}$ can be expanded in the unitary gauge as 
\begin{align}
    \mathcal{H}(x)=\frac{1}{\sqrt{2}}\begin{pmatrix}
    0\\ H + h(x) 
    \end{pmatrix}.
\end{align}
where only the real part of the neutral component has a constant background value $H$ and the physical higgs boson is denoted by $h$. At the minimum adopted by the universe we have $H= v_{H}$, where $v_{H}$ is the higgs vev. The EW symmetry is broken in the early universe if the scalar effective potential has no minimum in which $H$ vanishes. A sufficient condition for EW SNR is the effective 
mass squared $m_H^2(T)$ of the higgs field being negative at the field space points where $H=0$, i.e.~
\begin{equation}
    m_H^2(T) = \left. \frac{\partial^2 V(T,H)}{\partial H^2} \right|_{H=0} < 0. \label{SNRcond}
\end{equation}
Here $V(T,H)$ denotes the finite-temperature effective potential evaluated using the traditional background field method~\cite{Jackiw:1974cv}, which could also be a function of additional background fields. At finite temperatures, $m_H^2(T)$ acquires large positive contributions from the SM fields coupled to it, leading to the usual expectation of EW symmmetry restoration within the SM. All these SM particles contribute positively to $m_H^2(T)$ because the fermion and gauge-boson contributions are quadratic in their Yukawa and gauge couplings to the higgs, respectively. For the same reason, the EW symmetry remains to be restored in many early universe models with extended EW sectors. On the other hand, new scalar fields can couple with negative couplings to the higgs field and yield large negative contributions to $m_H^2(T)$. 

Consider, for instance, the simplest case where a real scalar field $S$ is coupled to the higgs doublet field $\mathcal{H}$ through a negative higgs-portal coupling $-\lambda_{HS}|\mathcal{H}|^2S^2/2$, which also couples $S$ to the SM thermal bath. Through this coupling, the thermal fluctuations of $S$ contribute  $\sim -\lambda_{HS}T^2$ to $m_H^2(T)$, which tend to push the higgs field away from the origin (i.e.~the field space points where the higgs background field vanish $H=0$). In order for one such scalar contribution to overcome the SM contributions while keeping the tree-level scalar potential  $V(H,S)$  bounded from below, the quartic self-coupling $\lambda_S$ of the $S$ field would need to be non-perturbatively large \cite{Meade:2018saz,Baldes:2018nel,Glioti:2018roy}. This led to the introduction of $O(100)$ scalars in Refs.~\cite{Meade:2018saz,Baldes:2018nel,Glioti:2018roy} in order to realize EW SNR in the early universe, while allowing for a perturbative treatment of the theory and keeping the tree-level potential bounded from below.

The preceding discussion assumes that the new scalar fields have no appreciable chemical potential, in which case the Bose-Einstein distribution corresponds to modes with energy $E_k\lesssim T$ having $O(1)$ occupation numbers. In the following we will consider more general momentum distributions. Schematically, the contribution to $m_H^2(T)$ from a scalar with arbitrary occupation numbers $f_k$ is proportional to $\int d^3k\; f_k/E_k$. This contribution is maximized for a given energy density $\rho\sim \int d^3k\; f_k E_k$ when the occupation number $f_k$ is concentrated in the infrared momentum modes, where the particle energy $E_k$ is minimized. Given its low entropy, it appears that a strongly-coupled field with an IR-concentrated momentum distribution would not last for a long time. However, such a configuration can be the favoured thermal-equilibrium state if there exists a non-zero chemical potential associated with some conservation law. In fact, when the chemical potential is close to a critical value, the configuration favoured by thermal equilibrium involves a large occupation of the ground state, i.e. a BEC. In that case, the contribution to $m_H^2(T)$ is maximized and, as we will show, this allows for a minimal realization of EW SNR with a single new scalar field.

In light of the above observation, we return to the real scalar singlet $S$ with a negative higgs-portal coupling example, now allowing it to develop a vev $v_S$. The effective mass squared of the higgs field at the origin is then given by
\begin{equation}
    \frac{m_H^2(T)}{T^2}\approx \kappa_{\rm SM}-\kappa_S
    \label{higgsTmass}
\end{equation}
where $\kappa_{\rm SM}$ and $\kappa_S=\lambda_{HS}v_S^2/(2T^2)$ are, respectively, the contributions from the SM thermal bath and the vev of $S$. At temperatures above the electroweak scale, the dominant contributions to $\kappa_{\rm SM}$ are \cite{Meade:2018saz}
\begin{equation}
    \kappa_{\rm SM}=\frac{y_t^2}{4}+\frac{3g^2+g^{\prime 2}}{16}+\frac{\lambda_H}{2}\approx 0.4\label{cSM}
\end{equation}
where $y_t$, $g$, $g^\prime$, and $\lambda_H$ are the top-Yukawa, $SU(2)_L$, $U(1)_Y$, and higgs self-quartic coupling, respectively. A sufficient condition for EW SNR is
\begin{equation}
    \frac{v_S}{T}\gtrsim 0.9\lambda_{HS}^{-1/2}\label{SNRreq}
\end{equation}
Thus, a single new scalar field that has a negative higgs-portal coupling and acquires a vev satisfying the above can reverse the EW symmetry restoring effect of the SM thermal bath. In the next section, we discuss a simple mechanism for sustaining such a large vev at high temperatures.

\section{Minimal Scalar Condensate Model}
\label{s:III}

We extend the SM with a complex scalar singlet $\phi$, playing the role of the $S$ field in the previous section. We assume that the tree-level scalar potential is invariant under a global $U(1)_\phi$ symmetry and include all allowed
renormalizable terms
\begin{align}
    V_{\rm tree}(\mathcal{H},\phi)=&-\mu_H^2|\mathcal{H}|^2+\lambda_H|\mathcal{H}|^4-\lambda_{H\phi}|\mathcal{H}|^2|\phi|^2\nonumber\\
    &+\mu_\phi^2|\phi|^2+\lambda_\phi|\phi|^4,
\label{tree-level}
\end{align}
where $\lambda_{H\phi}$, $\lambda_\phi$, and $\mu_\phi^2$ are all positive. This model has been studied extensively in connection to dark matter, baryogenesis, and gravitational wave production through a strong first order phase transition (for a review, see e.g. \cite{Lebedev:2021xey}). Unlike these previous studies, we assume that the universe has a pre-established large net conserved charge density $n_Q$ associated to the global $U(1)_\phi$ symmetry under which $\phi$ transforms. Such a charge density implies that there is an asymmetry $n_\phi-n_{\phi^\dagger}=n_Q$ in the number density of $\phi$ particle and antiparticle, denoted as $n_\phi$ and $n_{\phi^\dagger}$ respectively. We do not specify the origin of such a large charge asymmetry, but concrete mechanisms have been proposed \cite{Affleck:1984fy,Delepine:2006rn,Hertzberg:2013mba}. Since both $n_Q$ and entropy density $s$ scale with scale factor $a$ as $a^{-3}$, it is convenient to take their expansion-invariant ratio as a free parameter
\begin{equation}
    \eta_Q=\frac{n_Q}{s}
\end{equation}

\subsection{Thermal Bose-Einstein condensate}
We begin by specifying the conditions under which a BEC develops as a thermal-equilibrium state. Any net charge density that is carried by the nonzero momentum excitations of the $\phi$ field,  $n_Q^{k\neq0}$, manifests itself as an asymmetry in the Bose-Einstein distributions for the particles and antiparticles due to the existence of a non-vanishing chemical potential $\mu$
\begin{equation}
    n_Q^{k\neq0}=\int \frac{d^3k}{(2\pi)^3}\left(\frac{1}{e^{(E_k-\mu)/T}-1}-\frac{1}{e^{(E_k+\mu)/T}-1}\right).
\end{equation}
For definiteness, we take $\mu$ and hence the charge density to be positive. A larger chemical potential $\mu$ corresponds to larger $n_Q^{k\neq0}$ at a given temperature $T$. In order to keep the $\phi$ particle occupation number $[e^{(E_k-\mu)/T}-1]^{-1}$ positive, the chemical potential $\mu$ must not exceed the effective mass of $\phi$, $m_\phi^{\rm eff}$. The upper limit of the chemical potential, namely $m_\phi^{\rm eff}$, corresponds to the maximum charge asymmetry that can be accommodated in particle and antiparticle excitations at a given temperature $T$~\cite{Mangano:2001ix,Fukuyama:2007sx}
\begin{align}
    \eta_Q^{\rm crit}&=\left.\frac{n_Q^{k\neq0}}{s}\right|_{\mu\rightarrow m_\phi^{\rm eff}}\nonumber\\
    &\sim  
    \begin{cases}
        \displaystyle \frac{15}{2\pi^2 g_*} \left(\frac{m_\phi^{\rm eff}}{T}\right), &T\gtrsim m_\phi^{\rm eff}\\
        \displaystyle \frac{45 \zeta(3/2)}{\pi^2 g_*} \left(\frac{m_\phi^{\rm eff}}{2 \pi T}\right)^{3/2}, &T\lesssim m_\phi^{\rm eff}
    \end{cases}\label{BECreq}
\end{align}
where $g_*$ is the effective number of relativistic degrees of freedom. 

When the charge asymmetry $\eta_Q$ is larger than $\eta_Q^{\rm crit}$ the excess charge must be stored in the ground state, which leads to the development of a high occupation number ground state, i.e.~a BEC. In the presence of a BEC, the total change density $n_Q$ can be decomposed into two parts: the charge density stored in the condensate of $k=0$ quanta, $n_Q^{\rm BEC}$, and the charge density carried by the $k\neq 0$ particle excitations, $n_Q^{k\neq0}$.

The large occupation of the ground-state quanta in the condensate makes it possible to treat $\phi$ as a homogeneous classical field $\left<\phi\right>$, which can be written in terms of its radial component $r$ and phase $\theta$ as $\left<\phi\right>=re^{i\theta}/\sqrt{2}$.
In this viewpoint, the charged BEC corresponds to $\phi$ having a non-zero radial component $r$ coherently rotating in the field space with an angular velocity $\dot{\theta}$ \cite{Co:2019wyp}. The conserved charge density $n_Q$ of the spinning $\phi$ is given by its field-space angular momentum, $n_Q=\dot{\theta}r^2$. The equations of motion for $r$ read
\begin{align}
    \ddot{r}+(3H+\Gamma)\dot{r}-\dot{\theta}^2r+\partial_rV=0,
    \label{eom}
\end{align}
Here, $H$ is the Hubble rate and $\Gamma$ accounts for dissipative effects from the interaction of $\phi$ with the near-thermal SM plasma.  

The $\phi$ and SM sector can thermalize if the higgs-mediated $\phi$-$\phi^\dagger$ pair annihilation, i.e. the slowest reaction, is in equilibrium. Assuming that the kinetic-equilibration of the $\phi$ field occurs much more quickly, this process is efficient if
\begin{equation}
    \frac{\Gamma_{\rm th}}{H}\sim \frac{\lambda_{H\phi}^2n_{\phi^\dagger}^{k\neq0}/T^2}{T^2/M_{\rm P}}\sim \frac{\lambda_{H\phi}^2M_{\rm P}}{T}\gg 1
\end{equation}
with $M_{\rm P}\equiv 2.4\times 10^{18}\GeV$ being the reduced Planck mass, which is always satisfied at temperatures above the electroweak scale in the parameter space of our interest.\footnote{It is also possible that the early stages of the relaxation toward thermal equilibrium involve processes whose timescales are different from and possibly longer than the one considered here. For example, the early dissipation of the scalar $\phi$ may proceed through non-perturbative effects \cite{Mukaida:2012qn,Mukaida:2013xxa,Moroi:2013tea,Tanin:2017bzm} if its initial radial oscillation amplitude is sufficiently large. However, this is highly dependent on the reheating scenario and we do not consider it here.} We have assumed in this estimate that all the particles involved are relativistic, in which case $n_{\phi^\dagger}^{k\neq0}\sim T^3$. Once thermal equilibrium is established, the $\phi$ field moves in the field space in a circular orbit determined by the balance between the centrifugal force and potential gradient, $\dot{\theta}^2r=\partial_rV$. This configuration is nothing but the previously discussed thermal-equilibrium BEC.

\subsection{Electroweak asymmetric early universe}
We now describe the cosmology of our model at temperatures above the electroweak scale. In this high-temperature regime, we assume both the higgs field and the complex singlet $\phi$ acquire vevs. We write the fluctuations around the constant background values as
\begin{align}
    \mathcal{H}=\frac{1}{\sqrt{2}}\begin{pmatrix}
    0\\H+h(x)
    \end{pmatrix},\quad    
    \phi=\frac{1}{\sqrt{2}}\left(r+\varphi(x)\right)e^{i\theta(x)},
    \label{fieldexpansions}
\end{align}
where we reserve the notation $H = v_{H}(T)$ and $r= v_{r}(T)$ for the physical vevs adopted by the universe in our model at a temperature $T$.

To get a sense of the general evolution of the universe, we start by discussing some estimates where we include only tree-level effects and thermal masses. In this approximation, the equilibrium radial expectation value $v_r/\sqrt{2}$ of the complex scalar $\phi$ is determined by the balance between the ``centrifugal force" $\dot{\theta}^2v_r$ and the potential gradient of the tree-level $\phi$ potential $\left.\partial_rV_{\rm tree}\right|_{r=v_r}$, which for sufficiently large $\eta_Q$ is dominated by the contribution from the quartic term $V_{\rm tree}\supset \lambda_\phi r^4/4$. Setting $\dot{r}=0$ and $\ddot{r}=0$ in Eq.~\eqref{eom}, and using $V_{\rm tree}\approx \lambda_\phi r^4/4$ and $n_Q=\dot{\theta}v_r^2=(2\pi^2/45)\eta_Q g_* T^3$, we find
\begin{equation}
    \frac{v_r}{T}\approx 0.76\eta_Q^{1/3}g_*^{1/3}\lambda_\phi^{-1/6}\label{roverT}
\end{equation}
Comparing Eq.~\eqref{roverT} with Eq.~\eqref{SNRreq}, we find a rough requirement for EW SNR 
\begin{equation}
    \eta_Q\gtrsim 1.6 g_*^{-1}\lambda_{H\phi}^{-3/2}\lambda_\phi^{1/2}. \label{SNRreqmodel}
\end{equation}
It can be easily checked that in the regime where EW SNR occurs, as defined by Eq.~\eqref{SNRreqmodel}, the quartic self-interaction indeed dominates the potential of $\phi$,\footnote{We found that in the parameter space where EW SNR occurs, the $\phi$ condensate can dominate the energy density of the universe by a factor of $\sim \lambda_\phi v_r^4/g_*T^4\sim \eta_Q^{4/3}\lambda_\phi^{1/3}g_*^{1/3}$. Since the energy density of $\phi$ is dominated by its quartic potential, which scales with the scale factor like $a^{-4}$, the universe would still be effectively radiation-dominated. } the BEC formation condition Eq.~\eqref{BECreq} is temperature-independent and easily satisfied, and we can consider all the charge to be stored in the BEC, i.e.~$n_Q\approx n_Q^{\rm BEC}\gg n_Q^{\rm k\neq0}$. Furthermore, the higgs vev is non-zero at high temperatures 
\begin{equation}
    v_H(T)^2 \approx v_H(0)^2+\left(\frac{\kappa_\phi-\kappa_{\rm SM}}{\lambda_H}\right)T^2
    \label{vevtemp}
\end{equation}
where $v_H(0)=246\GeV$, $\lambda_H=0.13$, $\kappa_\phi\approx \lambda_{H\phi}(v_r/T)^2/2$ with $v_r/T$ shown in Eq.~\eqref{roverT}, $\kappa_{\rm SM}\approx 0.4$ as found in Eq.~\eqref{cSM}, and here $\kappa_\phi>\kappa_{\rm SM}$. In the parameter space of our interest, from Eq.~(\ref{vevtemp}) we observe that $v_H/T$ is at most $O(1)$ and mildly temperature dependent. The latter implies that the higgs vev scales linearly with temperature, meaning that the SM particles coupled to the higgs can be considerable heavier in the early universe, though this change is at most $O(1)$ of their thermal masses. 

To account for a more complete description of the phenomenon of EW SNR, the resummation of a certain set of higher order diagrams, the so-called daisy diagrams, should be included~\cite{Gross:1980br, PhysRevD.45.4695, PhysRevD.47.3546}.\footnote{Recent computations of the characteristics of first-order phase transitions that go beyond the usual daisy-resummed approach have been performed, for instance, in Refs.~\cite{Croon:2020cgk, Schicho:2022wty}, where it was shown that two-loop contributions to the effective potential can be sizeable. We leave a discussion including such improvements for future work.} Furthermore, below the electroweak scale, the change in $g_*$, the zero-temperature masses of the SM particles and of the scalar field $\phi$ must be taken into account, and moreover the $\phi$ scalar potential is no longer dominated by the quartic self-interaction. Finally, the Coleman-Weinberg potential~\cite{Coleman:1973jx}, which accounts for the one-loop radiative corrections at zero-temperature should also be considered. In the next section, we explain how we include all these effects in our numerical analysis and in our plots, and show that they do not affect our conclusions qualitatively.

\subsection{Viable parameter space and numerical analysis}
\label{s:C}

The finite-temperature effective potential for the background fields $H$ and $r$, as defined in Eq.~\eqref{fieldexpansions}, at one-loop, is given by
\begin{equation}
    V(H,r,T) = V_{\rm tree} + V_{Q} +  V_{\rm CW} + V_{\rm CT} + V_{T} + V_{\rm daisy},
    \label{effPOT}
\end{equation}
where the tree-level scalar potential $V_{\rm tree}$ reads
\begin{align}
    V_{\rm tree}=&-\frac{1}{2}\mu_H^2H^2+\frac{\lambda_H}{4}H^4-\frac{\lambda_{H\phi}}{4}H^2r^2\nonumber\\
    &+\frac{1}{2}\mu_\phi^2r^2+\frac{\lambda_\phi}{4}r^4
\end{align}
$V_Q$ is the angular part of the kinetic energy of $\phi$, $\dot{\theta}^2r^2/2$, which owing to the conservation of charge $n_Q=\dot{\theta}r^2$, acts like an effective potential for $r$
\begin{equation}
    V_Q=\frac{n_Q^2}{2r^2}
\end{equation}
Note that $V_Q$ ensures that $r>0$. $V_{\rm CW}$ represents the one-loop zero-temperature radiative corrections to the scalar potential in the form of the Coleman-Weinberg potential, $V_{\rm CT}$ contains UV-finite counterterm contributions, $V_{T}$ includes the one-loop temperature-dependent corrections to the scalar potential, and finally $V_{\rm daisy}$ accounts for the resummation of the daisy diagrams. The expressions for $V_{\rm CW}$, $V_{\rm CT}$, $V_{T}$, and $V_{\rm daisy}$ can be found in Appendix~\ref{appendix:Veff}.

In our numerical analysis, we choose a set of input parameters which include the higgs portal coupling $-\lambda_{H\phi}$, the singlet self-coupling $\lambda_\phi$, the singlet charge asymmetry per unit entropy $\eta_Q$, and the present-epoch
physical mass of the extra scalar  $m_{\phi}$,
\begin{equation}
    \lambda_{H\phi} \, , \lambda_\phi \,, m_{\phi} \,, \eta_Q.
\end{equation}
To ensure perturbativity and the boundedness of $V_{\rm tree}$, we require \cite{Costa:2014qga}
\begin{equation}\label{eq:breakingcond}
\begin{gathered}
    \lambda_{H}<4\pi,~\lambda_\phi <4\pi,~\lambda_{H\phi}<8\pi,\\
    \Lambda_{\pm} \equiv 6\lambda_H+4\lambda_\phi \pm \sqrt{(4 \lambda_\phi)^2+(6\lambda_H+4\lambda_\phi)^2} < 16\pi,\\
    4\lambda_H\lambda_\phi>\lambda_{H\phi}^2, \\
    \lambda_{H}, \lambda_{\phi} > 0.
\end{gathered}
\end{equation}
We look for the extrema of the effective potential where $H=0$. In order to asses whether a field-space extremum is a minimum, we calculated the principal minors of the Hessian matrix of the scalar potential shown in Eq.~\eqref{effPOT}. The EW symmetry could be restored if both minors are positive, otherwise the phenomenon of EW SNR would take place.

\begin{figure*}
\centering
\includegraphics[width=0.75\textwidth]{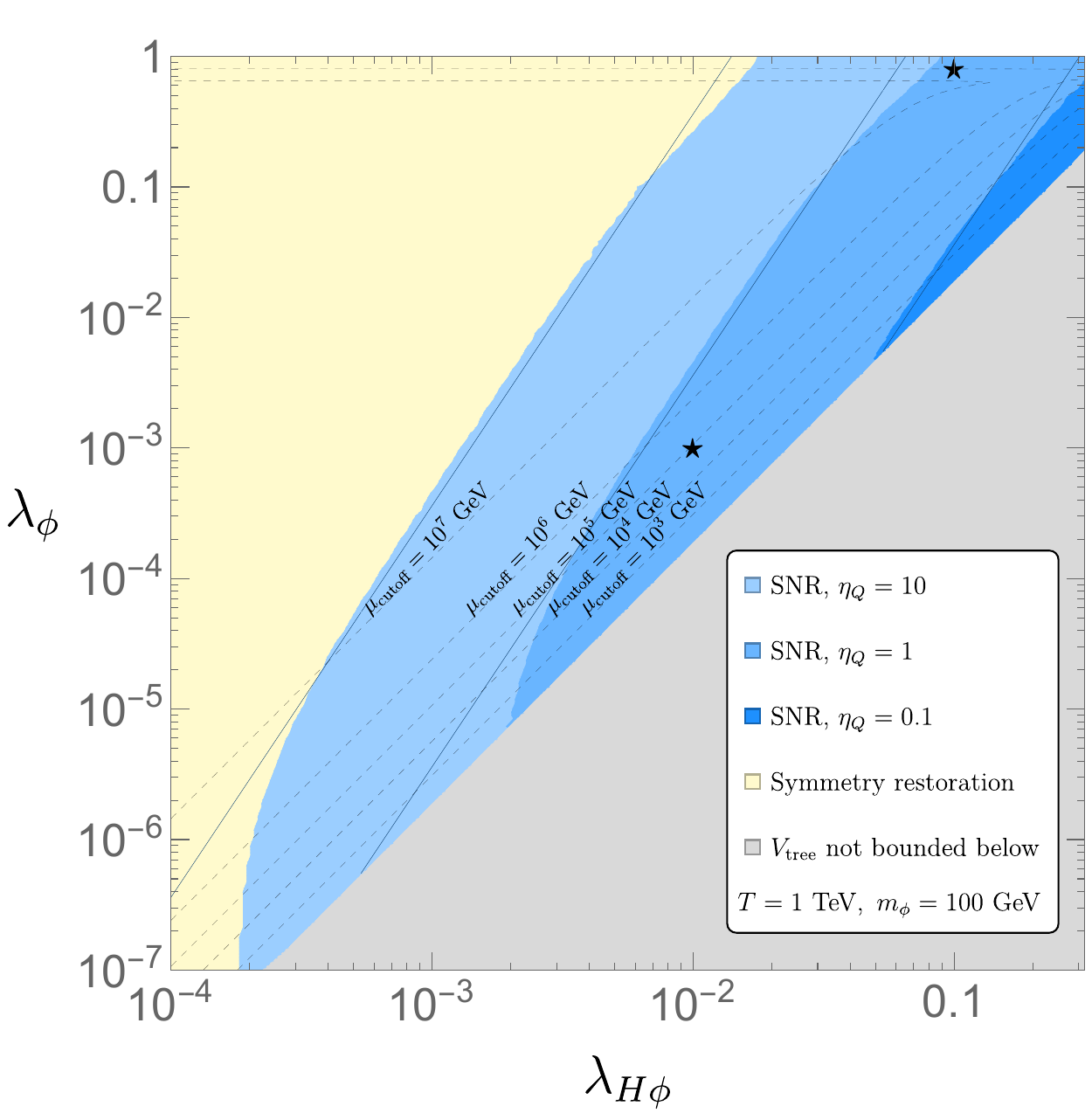}
\caption{$(\lambda_{H\phi},\lambda_\phi)$ parameter space for $\eta_Q=0.1,~1,~\textrm{and}~10$. The blue regions are the parameter space zones in which the EW symmetry is not restored at $T=1~\TeV$ for different values of $\eta_Q$, the yellow region features the possible restoration of the EW symmetry at the same temperature, and the grey region is where the tree-level potential is not bounded from below. The dark blue lines show the analytic approximation from Eq.~\eqref{SNRreqmodel} for the corresponding values of $\eta_Q$. The dashed lines are the cutoff energy scales derived from either perturbativity or boundedness constraints Eq.~\eqref{eq:breakingcond}. The RGE running of couplings for the parameter points marked with stars are shown in Fig.~\ref{fig:RGE}. }
\label{fig:parameterspace}
\end{figure*}

Our results are summarized in Fig.~\ref{fig:parameterspace}, which shows the parameter space where EW SNR is achieved at $T=1~\TeV$. 
To illustrate the viable parameter space, we assume representative values of the charge asymmetry $\eta_Q$ and the scalar singlet mass $m_{\phi}$, and scan over the couplings $\lambda_{H\phi}$ and $\lambda_{\phi}$. Charge asymmetries as large as $\eta_Q\sim 1$ are a natural outcome of the Affleck-Dine mechanism \cite{Affleck:1984fy,Linde:1985gh}. Therefore, we fixed $\eta_Q = 0.1, 1,~\textrm{and}~10$ in our parameter scan. As per Eq.~\eqref{BECreq}, the $\phi$ condensate automatically disappears at low temperatures $T\ll m_\phi$ where $\eta_Q^{\rm crit}\gg 1$ and so we assume $\left<\phi\right>=0$ when considering collider constraints. We set the singlet mass at $m_{\phi} = 100 \GeV$, given that a singlet mass $m_{\phi}$ larger than $m_h/2$ ensures agreement with the measured properties of the detected SM-like higgs boson at $125 \GeV$~\cite{CMS:2022dwd,ATLAS-CONF-2022-050}, since the decay of the higgs boson into a pair of additional scalars is forbidden. The blue regions in Fig.~\ref{fig:parameterspace} demonstrate the zones where the EW symmetry is not restored at $T=1 \TeV$ for the three different values of $\eta_Q$, the yellow region indicates where the EW symmetry could be restored at that temperature, and the grey region features a tree-level potential not bounded below. The dashed lines depict the cutoff energy scales $\mu_{\rm cutoff}$ above which one of the conditions in Eq.~\eqref{eq:breakingcond} is not satisfied when the running couplings are inserted therein. The running couplings were obtained by solving the one-loop renormalization group equations (RGE) in the $\overline{\mathrm{MS}}$ renormalization scheme shown in Appendix~\ref{RGEappendix}. On one hand, the dashed lines parallel to the boundary between the grey and the blue regions indicate the cutoff energy scales above which tree-level boundedness from below with inserted one-loop running couplings is not fulfilled anymore. On the other hand, the dashed horizontal lines show the values of $\mu_{\rm cutoff}$ for which perturbativity breaks down.

\begin{figure}
\centering
\includegraphics[width=0.48\textwidth]{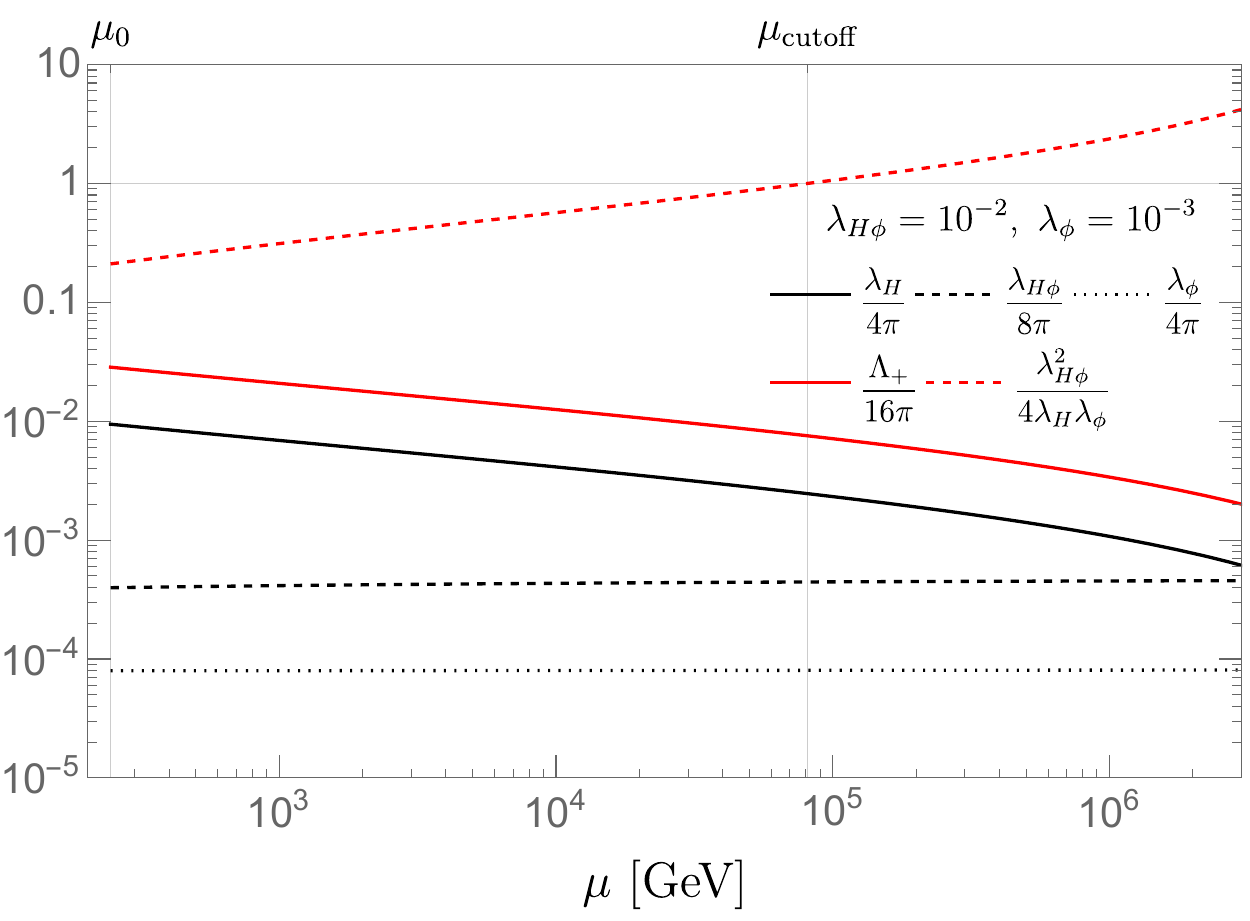}\\
\vspace{0.5 cm}
\includegraphics[width=0.48\textwidth]{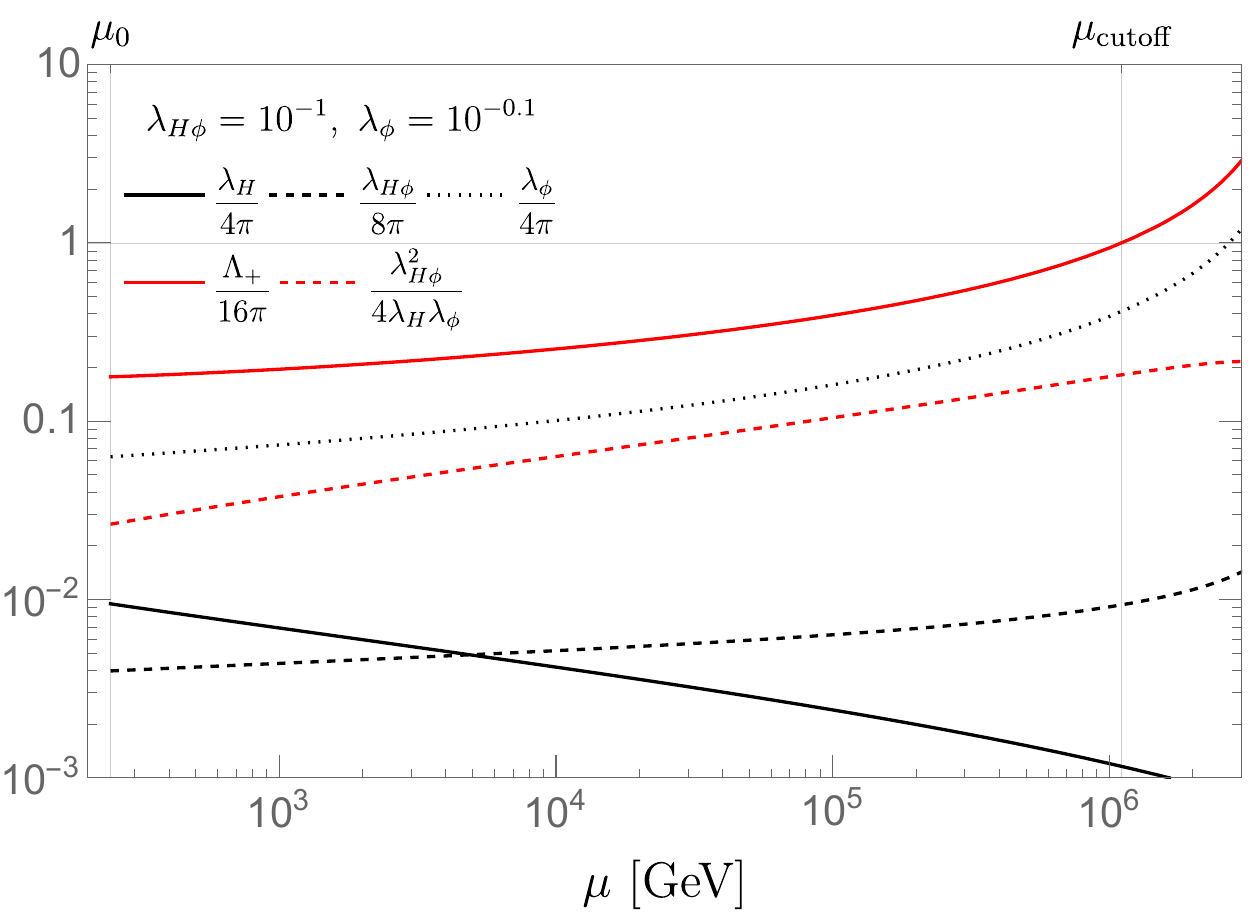}
\caption{The RG flow of couplings for $(\lambda_{H\phi},\lambda_\phi)$ at the reference scale $\mu_0=246~\GeV$ equal to $(10^{-2},10^{-3})$ and $(0.1,10^{-0.1})$, which are marked in Fig.~\ref{fig:parameterspace} with stars. The cutoff scale $\mu_\textrm{cutoff}$ is given by the lowest energy scale at which any of the curves exceeds 1. $\mu_\textrm{cutoff}$ is given by the breakdown of tree-level boundedness (perturbativity) for the upper (lower) plot.}
\label{fig:RGE}
\end{figure}

The running of the quartic couplings for two benchmark scenarios is shown in Fig.~\ref{fig:RGE}, whose parameter points are marked in Fig.~\ref{fig:parameterspace} with stars. The upper plot in Fig.~\ref{fig:RGE} corresponds to a representative case where $\mu_{\rm cutoff}$ is set by the tree-level boundedness from below requirements, while the lower plot exemplifies a situation where $\mu_{\rm cutoff}$ is given by the breakdown of perturbativity. In most cases, the absolute value of $\lambda_{H \phi}$ increases with the energy scale due to the contribution from the top quark to its beta function. Furthermore, $\lambda_{H}$ decreases with the renormalization energy scale. According to equation~\eqref{vevtemp}, these facts make it easier to achieve EW SNR for temperatures much higher than $1 \TeV$ when one considers the RG improved effective potential, which is needed to minimize the renormalization scale dependence. In this paper, we demonstrated the realization of EW SNR at $T=1 \TeV$, where the condition to achieve this phenomenon is expected to be stronger. At $T=1 \TeV$, the effective potential is expected to be only mildly renormalization-scale dependent.\footnote{We numerically checked that at $T=1 \TeV$, $v_{r}$ is of order $O(1 \TeV)$ at most.} Therefore, we did not consider the RGE improved effective potential in our analysis.

\subsection{Late-time charge washout}

The low-temperature ($T \lesssim 100 \GeV $) part of the cosmological history can be told in any number of ways without affecting the previous, high-temperature ($T \gtrsim 100 \GeV $) discussion. Since we are mainly concerned with achieving EW SNR at high temperatures, we keep the discussion here mostly qualitative. In order not to affect Big Bang nucleosynthesis (BBN), the universe must be dominated by SM radiation by  $T\sim 10\MeV$ \cite{Cyburt:2015mya}. This requires a process that depletes the energy density of the $\phi$ field taking place in the temperature range $10\MeV\lesssim T\lesssim 100\GeV$. Here, we give one concrete example of such a process that is consistent with current experimental and observational data without introducing new constraints to the high-temperature cosmological evolution.

For illustration purposes, we set
\begin{equation}
    \eta_Q\sim 0.1,\quad \lambda_{H\phi}\sim\lambda_\phi\sim 1,\quad m_{\phi}=100\GeV
\end{equation}
From Eq.~\eqref{BECreq}, we see that the BEC evaporates away at $T\sim 10 \GeV$. Below that temperature it suffices to describe $\phi$ as a particle-antiparticle gas. By that point, $\phi$ is non-relativistic, the universe is $\phi$ dominated (i.e.~matter dominated), and the antiparticles $\phi^\dagger$ have mostly annihilated away due to Boltzmann suppression. Note that $\eta_Q$ and $g_*$ always appear in conjunction in our calculations in this section and $\eta_Q g_*\sim 1$ at $T\lesssim 10\GeV$. For simplicity, we will thus omit them, together with $\lambda_{H\phi}\sim\lambda_\phi\sim 1$, in the estimates below.

In order to deplete the charge asymmetry, and hence the energy density of the $\phi$ field, we introduce the following $U(1)_\phi$ breaking terms
\begin{equation}
    V=\frac{1}{2}\Delta^2(\phi^2+\phi^{\dagger 2})
\end{equation}
where $\Delta$ can be made real without loss of generality. For a sufficiently small $\Delta$, these terms become important only at late times, and then they cause oscillations between the $+1$ and $-1$ charge eigenstates, namely $\phi$ and $\phi^\dagger$, respectively. Such oscillations repopulate the otherwise Boltzmann-suppressed number density of $\phi^\dagger$ particles, thus enabling the depletion of $\phi$ energy density through $\phi$-$\phi^\dagger$ annihilations \cite{Buckley:2011ye, Cirelli:2011ac,Tulin:2012re}. 

Before we describe the depletion process and derive the relic $\phi$ abundance, let us first outline the key processes involved. A freely propagating $\phi$ particle would oscillate into $\phi^\dagger$ with a probability $P_{\phi\rightarrow \phi^\dagger}\approx \sin^2(\Delta^2 t/m_\phi)$ over a time $t$. However, such oscillations are interrupted because the abundant $\phi$ particles are constantly colliding off one another, $\phi\phi\rightarrow\phi\phi$, with the rate $\Gamma_{\rm scat}\sim n_\phi T/m_\phi^3$. Since each of these collisions projects the participating particles into one of the charge eigenstates, the oscillations occur effectively only in the mean free time $\Gamma_{\rm scat}^{-1}$ between two collisions. Hence, whenever two quanta collide, there is a probability $P_{\phi\rightarrow\phi^\dagger}\sim \left(\Delta^2/m_\phi \Gamma_{\rm scat}\right)^2$  that one of them is projected into $\phi^\dagger$. The rate at which $n_\phi$ is converted to $n_{\phi^\dagger}$ through oscillations is thus given by
\begin{equation}
    \dot{n}_{\rm osc}\sim n_\phi \Gamma_{\rm scat}P_{\phi\rightarrow\phi^\dagger}\sim \Delta^4\frac{m_\phi}{T}   
\end{equation}
which incidentally depends on neither $n_\phi$ nor $n_{\phi^\dagger}$. Furthermore, a $\phi$-$\phi^\dagger$ pair can annihilate into SM states via higgs-mediated processes with the rate per unit volume $\sim \left<\sigma v\right>_{\rm ann} n_\phi n_{\phi^\dagger}$. Finally, both the particle and antiparticle number density are diluted by the expansion of the universe. These effects are summed up by the following schematic Boltzmann equations
\begin{align}
    \dot{n}_\phi&\sim -\dot{n}_{\rm osc}-\left<\sigma v\right>_{\rm ann}n_\phi n_{\phi^\dagger}-3Hn_\phi\\
    \dot{n}_{\phi^\dagger}&\sim +\dot{n}_{\rm osc}-\left<\sigma v\right>_{\rm ann}n_\phi n_{\phi^\dagger}-3Hn_{\phi^\dagger}
\end{align}
where $H$ is the Hubble rate, and we assumed $n_\phi\gtrsim n_{\phi^\dagger}$ and correspondingly neglected the $\phi^\dagger\rightarrow \phi$ oscillation. For our choice of parameters the annihilation cross section is $\left<\sigma v\right>_{\rm ann}\approx 2.0\times 10^{-9}\GeV^{-2}$.\footnote{This annihilation cross section $\left<\sigma v\right>_{\rm ann}\approx 2.0\times 10^{-9}\GeV^{-2}$ was computed with the analytical formula $\left<\sigma_{\rm ann}v\right>=[\lambda_{ H\phi}^2v_H^2/2m_\phi(m_h^2-4m_\phi^2)^2]\left.\Gamma_h\right|_{m_h\rightarrow 2m_\phi}$ from \cite{Burgess:2000yq} using the higgs decay width $\Gamma_h$ from \cite{LHCHiggsCrossSectionWorkingGroup:2011wcg}.}

Our choice of parameters is such that initially oscillations and annihilations dictate the evolution of $n_{\phi^\dagger}$, while $n_\phi$ is barely affected by the oscillations and simply dilutes as the universe expands. In a timescale $\sim (\left<\sigma v\right>_{\rm ann} n_\phi)^{-1}$ much less than a Hubble time, during which $n_\phi$ is approximately constant, $n_{\phi^\dagger}$ settles into a quasi-equilibrium where the oscillation and annihilation rates per unit volume balance, $\dot{n}_{\rm osc}\sim \left<\sigma v\right>_{\rm ann} n_\phi n_{\phi^\dagger}$. In several Hubble times that follow, this balance is maintained while $\dot{n}_{\rm osc}\propto T^{-1}$ increases and $3H n_\phi$ decreases. Eventually, the two meet $\dot{n}_{\rm osc}\sim 3H n_\phi$ when the Hubble rate is
\begin{equation}
    H_{\rm osc}\sim \left(\frac{\Delta^{12}m_\phi^{7}}{M_{\rm P}^8}\right)^{1/11} \label{Hosc}
\end{equation}
where we have assumed that the universe is $\phi$-dominated when this happens, $H^2\sim T^3m_\phi/M_{\rm P}^2$. At that point, 
$n_\phi\sim n_{\phi^{\dagger}}$ and the $\phi$ field is approximately symmetrized.

Henceforth, the annihilation rate per unit volume is given by the usual symmetric thermal freeze-out formula, and the density of $\phi$ at freeze-out is given by $n_\phi^{\rm FO}\sim H_{\rm FO}/\left<\sigma v\right>_{\rm ann}$, where $H_{\rm FO}$ is the Hubble rate when $\phi$ freezes out.\footnote{While the universe is initially dominated by $\phi$ (i.e. matter dominated), its energy density is quickly transferred to the SM sector as it annihilates away. Hence, the later part of the depletion process that determines the relic abundance of $\phi$ takes place during SM radiation domination, similar to the standard thermal freeze-out.} The abundance of $\phi$ today is given by
\begin{equation}
    \Omega_\phi^{\rm FO}\sim \frac{1}{\rho_{\rm crit,0}} m_\phi n_\phi^{\rm FO}\frac{T_0^3}{T_{\rm FO}^3}
    \label{OmegaFO}
\end{equation}
where $T_{\rm FO}\sim \sqrt{H_{\rm FO}M_{\rm P}}$ is the Standard Model temperature at $\phi$ freeze-out and $\rho_{\rm crit,0}$ and $T_0$ are the (fixed) critical density and temperature of the Standard Model today. In the standard symmetric freeze-out scenario, the freeze-out happens when $T_{\rm FO}\sim m_\phi/25$ at which $H_{\rm FO}\sim 10^{-3}m_\phi^2/M_{\rm P}$. To keep the standard freeze-out prediction unaltered, we assume that the $H_{\rm osc}$ found in Eq.~\eqref{Hosc} is above $H_{\rm FO}$, which amounts to requiring $\Delta\gtrsim 10\keV$ for the current choice of parameters. We also require $\Delta\lesssim 10\MeV$ such that $H_{\rm osc}\lesssim H(T\sim 100\GeV)$ to avoid complicating the EW SNR discussion in the previous sections.\footnote{Since the $\Delta^2$ Lagrangian terms are the only ones that break the global $U(1)_\phi$ symmetry at low energies, loop corrections to $\Delta^2$ are proportional to itself, thus making its small value technically natural \cite{Cui:2009xq}.}

The low-temperature phenomenology of our $\phi$ field has some resemblance with that of the minimal singlet scalar dark matter (complex or real) coupled to the SM via the higgs portal \cite{Burgess:2000yq, PhysRevD.88.055025,Lebedev:2021xey}. It is known that such models are essentially ruled out if the scalars are populated via symmetric thermal freeze-out, even when they are allowed to be a small part of the dark matter. The combined constraints from collider experiments, agreement with the measured dark matter relic abundance~\cite{Planck:2018vyg}, bounds from direct detection of dark matter, and perturbativity of $\lambda_{H\phi}$ leave no viable parameter space apart from a narrow mass regime around half of the higgs boson mass where the annihilation rate is resonantly enhanced \cite{Escudero:2016gzx,Lebedev:2021xey,Biekotter:2022ckj}. Since the lowest $\phi$ abundance predicted by the ``initially asymmetric" freeze-out mechanism described above matches that of the standard symmetric freeze-out for a given mass $m_\phi$, our scenario requires a process that further dilutes the abundance of $\phi$. A new field $\psi$ with mass $m_\psi$ that comes to dominate the universe and decays into the Standard Model after the freeze-out of $\phi$ can increase the Standard Model entropy by a factor of $\zeta\sim m_\psi/T_{\rm RH}\lesssim 10^3$, where $T_{\rm RH}$ is the Standard Model reheating temperature right after the $\psi$ decay, leading to a suppression in $\Omega_\phi$ by the same factor $\zeta$ \cite{Randall:2015xza, Bramante:2017obj,Chanda:2019xyl,Evans:2019jcs}. The strongest dilution $\zeta\sim 10^3$ corresponds to $m_\psi\sim m_\phi/10\sim 10\GeV$ ($\psi$ domination occurring after the earliest $\phi$ freeze-out) and $T_{\rm RH}\sim 10\MeV$ (lowest reheating temperature compatible with BBN). The lower-than-expected relic abundance $\Omega_\phi\lesssim \Omega_\phi^{\rm FO}$, where $\Omega_\phi^{\rm FO}$ is the relic abundance of $\phi$ in the standard freeze-out scenario found in Eq.~\eqref{OmegaFO}, simultaneously avoids dark matter overproduction and relaxes the constraints from direct detection, whose rate scales as $R_{\rm DD}\propto \Omega_\phi/\Omega_\phi^{\rm FO}$ \cite{Craig:2014lda}. Furthermore, since the singlet vev vanishes ($v_{r}(0)=0$) when the BEC evaporates below the electroweak scale, as far as collider searches are concerned there is no mixing between the higgs field $h$ and the scalar $\phi$, leaving the couplings of the higgs boson to SM particles unchanged with respect to the SM. Thus, collider bounds on our model are currently essentially non-existent~\cite{Curtin:2014jma}.

\section{Discussion and Conclusion}
\label{s:IV}

Current observations do not preclude the possibility that the electroweak symmetry remains broken at temperatures above the electroweak scale in the early universe. In such electroweak symmetry non-restoration (EW SNR) scenarios, the higgs vev remains non-zero and typically grows with temperature in the early universe. This can lead to early universe cosmological histories radically different from what is commonly assumed. Given its wide-ranging phenomenological consequences~\cite{Meade:2018saz}, the possibility of EW SNR is an interesting scenario.

EW SNR requires negative higgs mass squared contributions at high temperatures to give the higgs field a vev. In most versions of EW SNR proposed so far, these arise from the thermal fluctuations of new fields coupled to the higgs field. The difficulties associated with ensuring the boundedness of the scalar potential and the perturbativity of quartic couplings in such scenarios are usually overcome by increasing the number of new fields. In this paper, we argue that EW SNR can be minimally realized if the higgs field is repelled from the origin by the large vev of a single field, instead of the thermal fluctuations of many fields.

To demonstrate this, we extend the higgs sector with a complex scalar singlet with a quartic self-interaction $\lambda_{\phi}|\phi|^4$ and introduce a negative higgs-portal coupling $-\lambda_{H\phi}|\mathcal{H}|^2|\phi|^2$ to yield EW SNR. If the singlet is in thermal equilibrium with a pre-established charge asymmetry $\eta_Q$ above a critical value, it acquires a large vev which then drives the higgs field away from the origin through the higgs-portal coupling, thus realizing EW SNR. Important constraints on $\eta_Q$, $\lambda_{H\phi}$, and $\lambda_\phi$ to achieve a viable scenario where EW SNR is realized come from tree-level perturbative unitarity and boundedness of the scalar potential from below. The renormalization-group running of quartic couplings triggers the breakdown of these two requirements at higher temperatures, but in general EW SNR can be realized up to a upper cutoff-temperature many orders of magnitude above the electroweak scale.

The high-temperature EW SNR scenario we are proposing is independent of any specific scenario below the electroweak scale. While constraints arising from low-temperature observables such as those from BBN, dark matter abundance, direct and indirect detection experiments, and colliders measurements may impose further constraints on the main model parameters  $\eta_Q$, $\lambda_{H\phi}$, and $\lambda_\phi$ that are relevant at high temperatures, these constraints depend on the cosmological evolution at temperatures below the EW scale. We found a simple low-temperature scenario where no new constraint is added to the main model parameters. In this scenario, we add a small term in the scalar potential, controlled by a mass scale $\Delta$, which softly breaks the $U(1)_\phi$ symmetry, leading to the depletion of the energy density of the singlet at some point below the electroweak scale and before BBN. The symmetry-breaking parameter $\Delta$ and the bare mass $m_{\phi}$ of the singlet can be appropriately tuned to meet various constraints without affecting the high-temperature realization of EW SNR.

The most promising way to probe the model under consideration at present colliders would be through the searches for invisible decays of the higgs boson, for a set-up in which the singlet field is lighter than half of the higgs boson mass.
The case in which the singlet state is heavier than half of the higgs boson mass could only be accessible to future colliders, such as a $100 \TeV$ hadron collider or a $1 \TeV$ electron-positron collider, by means of precision measurements of the triple higgs boson self-coupling and the $Zh$ production cross section, and only for the largest values of $\lambda_{H\phi}$ explored in our analysis~\cite{Curtin:2014jma}. The specific depletion scenario we considered predict WIMP-like relics which can be tuned to make up the entirety of the dark matter or a sub-component of it. If the relic abundance of the $\phi$ field makes up a significant part of the dark matter relic abundance, we could probe this scenario in present and future direct and indirect detection experiments. The novel freeze-out mechanism proposed here could deserve special attention from the viewpoint of dark matter model building. We leave the detailed study of this scenario for future work

It would be interesting to see if EW SNR could be realized in other models involving scalar condensates. The $U(1)$-symmetric scalar condensate model considered here can be generalized to fields that respect wider global symmetries \cite{Li:2001xaa,Moore:2015adu}. The required chemical potential to support a Bose-Einstein condensate may also arise from the departure from thermal equilibrium, in which case the IR-dominated distribution function may be achieved as an initial condition through the decay a non-relativistic field, e.g. the inflaton, and kept from cascading toward the UV by suppressing the thermalization rates with the Standard Model plasma \cite{Tenkanen:2016jic}. Other ways to generate a non-zero scalar expectation value in the early universe include coupling the scalar field to fermions \cite{Batell:2021ofv} and introducing a non-minimal coupling to gravity \cite{Cosme:2018nly}.

\section*{Acknowledgments}
We thank Iason Baldes, Thomas Biekötter, Raymond Co, Pierre Fayet, Sven Heinemeyer, Zhen Liu, Oleksii Matsedonskyi, Kalliopi Petraki, Harikrishnan Ramani, G\'eraldine Servant, and Yikun Wang for useful discussions. JHC is supported by the NSF grant PHY-1914731, the Maryland Center for Fundamental Physics, and the JHU Joint Postdoc Fund. MOOR is supported by the European Union’s Horizon 2020 research and innovation programme under grant agreement No 101002846, ERC CoG ``CosmoChart". The work of MOOR was also supported by the Deutsche Forschungsgemeinschaft under Germany’s Excellence Strategy EXC2121 “Quantum Universe” - 390833306. This research was supported in part by Perimeter Institute for Theoretical Physics. Research at Perimeter Institute is supported by the Government of Canada through the Department of Innovation, Science and Economic Development and by the Province of Ontario through the Ministry of Research and Innovation.

\appendix

\section{Scalar Effective Potential at One-loop}
\label{appendix:Veff}
In the following we present the different pieces that constitute the finite-temperature effective potential $V$ (see Eq.~\eqref{effPOT}). $V_{\rm CW}$ is given in the $\overline{\mathrm{MS}}$ renormalization scheme and in the Landau gauge by\footnote{The Coleman-Weinberg potential introduces gauge dependencies, however it appears only in subleading terms \cite{Patel:2011th,Garny:2012cg}.}
\begin{equation}
V_{\rm CW} = \sum_{j}\frac{n_{j}}{64\pi^{2}}(-1)^{2s_{j}} \,
m_{j}^{4}(H,r)
\left[\ln\left(\frac{|m_{j}(H,r)^{2}|}{\mu_0^{2}}\right)-c_{j}\right],
\label{CW_potential}
\end{equation}
where $m_{j}(H,r)$ are the field-dependent tree-level masses of all the particle species, $s_{j}$ the particle spins, $n_{j}$ the corresponding numbers of degrees of freedom, and $\mu_0$ is the reference renormalization scale. The constants $c_{j}$ arise from the $\overline{\mathrm{MS}}$ renormalization prescription,
with $c_j=5/6$ for gauge bosons and $c_{j}=3/2$ for scalars and fermions. The sum in~Eq.~\eqref{CW_potential} runs over the two scalar mass eigenstates, the three goldstone bosons $G_{i}$, the longitudinal and transversal gauge bosons,
$V_{L}=\{Z_{L},\,W_{L}^{+},\,W_{L}^{-}\}$ and $V_{T}=\{Z_{T},\,W_{T}^{+},\,W_{T}^{-}\}$ and the SM quarks $q$, and
leptons $\ell$. The degrees of freedom $n_{j}$ for the species of each type are $n_{h}=1$, $n_{\phi}=2$, $n_{V_{T}}=2$, $n_{G_{i}}=1$, $n_{V_{L}}=1$, $n_{q}=12$ and $n_{\ell}=4$. The mass matrix for the scalar degrees of freedom $\{h(x),\varphi(x)\}$ obtained from $V_{\rm tree} + V_{Q}$ is
\begin{equation}
    M^{2} = \left(\begin{array}{cc}
- \mu_{H}^{2} + 3\lambda_{H}H^{2} -\frac{\lambda_{H\phi}}{2}r^{2} & -\lambda_{H\phi}H r\\ 
-\lambda_{H\phi}H r & \mu_{\phi}^{2}+6\lambda_{\phi}r^{2}-\frac{\lambda_{H\phi}}{2}H^2
\end{array}\right).
\label{mass_matrix}
\end{equation}
Moreover, the three goldstone bosons $G_{i}$ ($i=1,2,3$) tree-level field-dependent masses read
\begin{equation}
    m_{G_{i}}^{2} = - \mu_{H}^{2} + \lambda_{H}H^{2} -\frac{1}{2}\lambda_{H\phi}r^{2}.
\end{equation}
Due to the singlet nature of the additional scalar $\phi$, the field-dependent masses of the gauge bosons and fermions depend only on $H$, and they are given by $m_{W}^{2}(H) = (g^{2}/4)H^2$, $m_{Z}^{2}(H) = (g^{2}+g'^{2})H^2/4$ and $m_{f}^{2}(H) = (y_{f}^{2}/2)H^{2}$, where $y_{f}$ is the higgs Yukawa coupling to the fermion $f$.

Additionally, we require the zero-temperature loop-corrected vacuum expectation values and scalar masses to be equal to their tree-level values, and we refer to this prescription as the ``on-shell'' (OS) renormalization. To achieve this,
we add a set of UV-finite counterterms $V_{\rm CT}$ to the effective potential, whose computation was done by following the methods described in~\cite{Biekotter:2021ysx, Basler:2016obg} and with the help of the public code \texttt{BSMPT}~\cite{Basler:2018cwe, Basler:2020nrq}.
At finite temperature $T$, the one-loop effective potential receives thermal corrections $V_T$, given by~\cite{Dolan:1973qd,Quiros:1999jp}
\begin{equation}
V_{T} =\sum_{j} \, \frac{n_{j} \, T^{4}}{2\pi^{2}}\,
J_{\pm}\left(\frac{m_{j}^2(H,r)}{T^{2}}\right)  ,
\label{thermal_potential}
\end{equation}
where the thermal integrals $J_{-}$ for bosons and $J_{+}$ for fermions 
are defined by
\begin{equation}
J_{\pm}(y^{2}) = \mp
\int_{0}^{\infty}dx\,x^{2}\,\log\left[1\pm\exp\left(-\sqrt{x^{2}+y^{2}}\right)\right]\, .
\label{thermalfunctions}
\end{equation}
Beside the degrees of freedom considered in~Eq.~\eqref{CW_potential}, the sum in~Eq.~\eqref{thermal_potential} includes the photon.  We used the implementation of $V_T$ of the public
code \texttt{CosmoTransitions}~\cite{Wainwright:2011kj}. Furthermore, we resummed the daisy diagrams through the Arnold-Espinosa method~\cite{PhysRevD.45.4695,PhysRevD.47.3546} (see also \cite{Croon:2020cgk}), which amounts to  adding the following additional contribution to the one-loop effective potential at finite temperature\footnote{We also included in the sum an overall factor $e^{-m_{k}(\varphi,r)/T}$ to avoid spurious contributions to the effective potential coming from $V_{\rm daisy}$.}
\begin{equation}
V_{\text{daisy}}=-\sum_{k}\frac{T}{12\pi}
\left\{ \left[m_k^2(H,r,T)\right]^{\frac{3}{2}} - \left[m_{k}^2(H,r,0)\right]^{\frac{3}{2}} \right\} \ 
\label{Daisy1resummation}
\end{equation}
where the sum in $k$ runs over the bosonic degrees of freedom, and
$m_k^2(H,r,T)$ and  $m_k^2(H,r,0)$ denote, respectively, their physical masses at finite temperature $T$ and at zero temperature~\cite{Carrington:1991hz}. The gauge boson thermal masses can be found in~\cite{Curtin:2014jma}. The physical masses at finite temperature $m_k^2(H,r, T)$ for the scalar degrees of freedom are obtained through the diagonalization of $M^{2}+ \text{diag}(\kappa_{\rm SM}, c_{\phi}) \, T^2$, where $\kappa_{\rm SM}$ is defined in Eq.~\eqref{cSM} and 
\begin{equation}
    c_{\phi} = \frac{3n_{Q}^{2}}{T^{2}}\frac{1}{r^{4}} -\frac{\lambda_{H\phi}}{6}+\frac{\lambda_{\phi}}{4}.
\end{equation}

\section{Beta functions}
\label{RGEappendix}

We computed the RGEs by using the public code \texttt{SARAH}~\cite{Schienbein:2018fsw, Staub:2013tta} to be
\begin{align}
    16 \pi^2\beta_{\lambda_{H}} =&  24\lambda_{H}^2 + \lambda_{H\phi}^2 -6y_{t}^4 + \frac{27}{200}g_{1}^{4}+\frac{9}{20}g_{1}^{2}g_{2}^{2}    \nonumber\\
    &+ \frac{9}{8}g_{2}^{4} + \lambda_{H}(-\frac{9}{5}g_{1}^{2}-9g_{2}^{2}+12y_{t}^{2}) \\ 
    16 \pi^2\beta_{\lambda_{H\phi}} =&-\frac{9}{10}g_{1}^{2}\lambda_{H\phi}-\frac{9}{2}g_{2}^{2}\lambda_{H\phi}+ 12 \lambda_{H}\lambda_{H\phi}  \\
    &- 4\lambda_{H\phi}^{2}+8\lambda_{H\phi}\lambda_{\phi}+6\lambda_{H\phi}y_{t}^{2}\nonumber\\
    16 \pi^2\beta_{\lambda_{\phi}} =&  2\lambda_{H\phi}^{2}+20\lambda_{\phi}^{2}\\
    16 \pi^2\beta_{\mu_{H}^{2}} =&  -\frac{9}{10}g_{1}^{2}\mu_{H}^{2} - \frac{9}{2}g_{2}^{2}\mu_{H}^{2}+12\lambda_{H}\mu_{H}^{2} \\
    &+2\lambda_{H\phi}m_{\phi}^{2}+ 6\mu_{H}^{2}y_{t}^{2}\nonumber\\
    16 \pi^2\beta_{m^{2}_{\phi}} =&  4 \lambda_{H\phi}\mu_{H}^{2}+8\lambda_{\phi}m^{2}_{\phi} 
\end{align}
with $y_t(\mu_0)=0.99$, $g(\mu_0)=0.65=g_2(\mu_0)$, $g'=0.35=\sqrt{3/5}g_1(\mu_0)$, $g_3(\mu_0)=1.4$, $\lambda_H(\mu_0)=0.13$, and $\mu_h^2=(88.4\GeV)^2$, where $\mu_0=246~\GeV$\cite{Heinemeyer:1998yj, ParticleDataGroup:2022pth,Bahl:2018qog}. We found agreement with the RGEs in~\cite{Costa:2014qga}. For solving these equations, the values of the parameters at the reference renormalization scale in the $\overline{\mathrm{MS}}$ renormalization scheme were needed. These values were obtained by performing a shift of the parameters defined at the reference renormalization scale in the OS renormalization scheme following~\cite{Biekotter:2021ysx}.

\newpage
\bibliography{references}
\bibliographystyle{apsrev4-1}

\end{document}